\documentclass{eas}
\usepackage{graphicx}
\newcommand{\zap}{ZAp}
\newcommand{\aap}{AAp}
\newcommand{\apj}{ApJ}
\newcommand{\apjs}{ApJS}

\newcommand{\teff}{T_{\mathrm{eff}}}
\newcommand{\logg}{\log g}
\newcommand{\feh}{\left[\mathrm{Fe}/\mathrm{H}\right]}
\newcommand{\ltaur}{\log\tau_{\mathrm{Ross}}}

\begin{document}

\title{The \textsc{Stagger}-grid: A Grid of 3D Stellar Atmosphere Models}
\address{Max-Planck-Institut f{\"u}r Astrophysik, Karl-Schwarzschild-Str. 1, 85741 Garching, Germany}
\address{Research School of Astronomy \& Astrophysics, Cotter Road, Weston ACT 2611, Australia}
\author{Z. Magic}\sameaddress{1,2}
\author{R. Collet}\sameaddress{2,1}
\author{M. Asplund}\sameaddress{2,1}
\begin{abstract}
Theoretical atmosphere models provide the basis for a variety of applications
in astronomy. In simplified one-dimensional (1D) atmosphere models,
convection is usually treated with the mixing length theory despite
its well-known insufficiency, therefore, the superadiabatic regime
is poorly rendered. Due to the increasing computational power over
the last decades, we are now capable to compute large grids of realistic
three-dimensional (3D) hydrodynamical model atmospheres with the realistic
treatment of the radiative transfer. We have computed the \textsc{Stagger}-grid,
a comprehensive grid of 3D atmosphere models for late-type stars.
In the presented contribution, we discuss initial results of the grid
by exploring global properties and mean stratifications of the 3D
models. Furthermore, we also depict the differences to classic 1D
atmosphere models.
\end{abstract}
\maketitle

\section{Introduction}

The observable spectrum of stars contains detailed information on
the physical conditions prevailing in stellar atmospheres. By the
comparison of observations with theoretical model predictions, one
can derive very detailed properties of stars, e.g., stellar parameters
or chemical compositions. Therefore, the accuracy of the conclusions
derived from observations hinges intimately on the quality of the
theoretical atmosphere models. Since the late 70's, the first grids
of 1D model atmospheres emerged and were continuously developed further
on (see Kurucz \cite{Kurucz:1979p4707}; Gustafsson et al. \cite{Gustafsson:1975p3743}), in particular,
major improvements were achieved in the microphysics leading to realistic
opacities (Castelli \& Kurucz \cite{Castelli:2004p4949}; Gustafsson et al. \cite{Gustafsson:2008p3814}).
However, in the recent decades the accuracy of the observations have
surpassed the predictive limits of simplified 1D models due to technical
developments, so that the drawbacks of 1D models are nowadays increasingly
apparent. In particular, modeling the convective energy transport
in the superadiabatic regime just below the optical surface of cool
stars depicts the most challenging aspect, and the mixing length theory
(MLT; B\"ohm-Vitense \cite{BohmVitense:1958p4822}) is still widely in use. 

Since convection is inherently a 3D non-local phenomenon, an appropriate
solution to this problems is solving the 3D hydrodynamical equations
for conservation of mass, momentum and energy coupled with a realistic
non-gray radiative transfer. Then the convective motions emerge self-consistently
from first principles leading to inhomogeneities and asymmetries in
the physical properties. Pioneering studies proved the superior predictive
capabilities of the 3D models in comparison with solar observations
(see Nordlund et al. \cite{Nordlund:2009p4109} for a comprehensive overview of 3D atmospheric convection).
We remark that this contribution states a summary of the published
work by Magic et al. \cite{Magic:2013}. For the interested reader we refer to
the latter, which contains more details and discussions.

\section{3D atmosphere models}

\begin{figure}
\includegraphics[width=88mm]{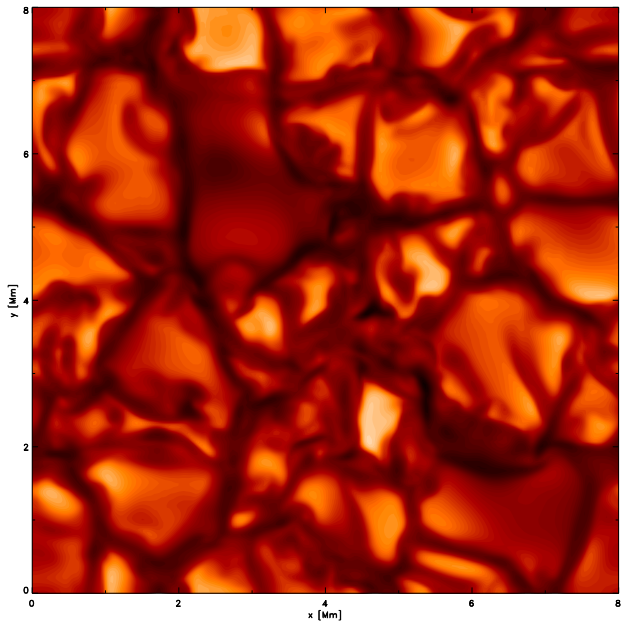}

\caption{\label{fig:solar_intensity}The bolometric intensity of the solar
simulation with the granulation pattern.}
\end{figure}
For the atmosphere simulations we applied the \textsc{Stagger}-code,
which is a state-of-the-art 3D radiative-hydrodynamic code. It solves
the time-dependent hydrodynamical equations for conservations of mass,
momentum and energy coupled with a realistic non-gray radiative transfer.
The radiative transfer is simplified with the so-called opacity binning
method (Nordlund \cite{Nordlund:1982p6697}; Skartlien \cite{Skartlien:2000p9857}), in order to
ease the computational load. We used the continuum and line opacities
mostly from the MARCS package (Gustafsson et al. \cite{Gustafsson:2008p3814}). Furthermore,
we employ a realistic equation of state provided by Mihalas et al. \cite{Mihalas:1988p20892},
which has been modified. The simulations of the so-called box-in-a-star
type are performed for a small statistical representative volume that
harbors approximately ten granules. The staggered Eulerian mesh is
horizontally periodic and vertically open. Furthermore, we note that
in Magic et al. \cite{magic:2013arXiv1307.3273M} we discuss the details of our
temporal and spatial averaging methods. 

In Fig. \ref{fig:solar_intensity}, we show the emergent intensity
map from our solar simulation with its typical stellar granulation
pattern, which comprises the bright granules arising from the bulk,
hot upflows and the dark intergranular lane given the by cold, narrow
downdrafts. Modern high-resolution solar observations cast a very
similar picture, which are almost indistinguishable from the theoretical
models.

\section{Global properties}

\begin{figure}
\includegraphics[width=88mm]{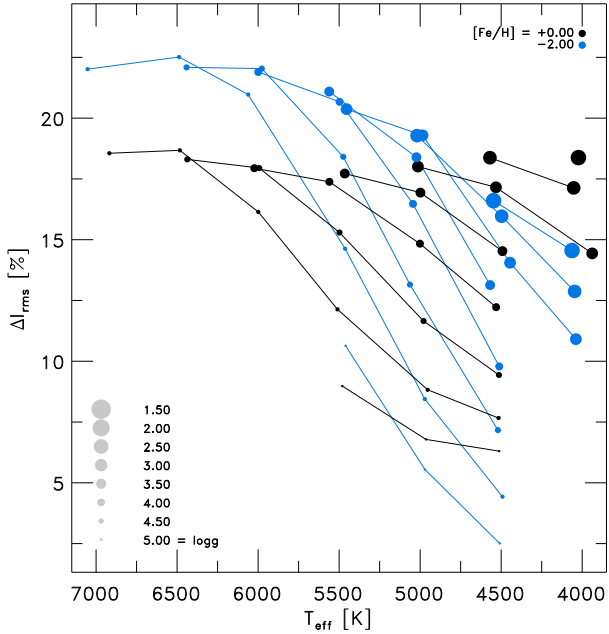}

\caption{\label{fig:intensity-contrast}The intensity contrast vs. $\teff$
for various stellar parameters.}
\end{figure}
The intensity contrast is usually quantified with the standard deviation
of the emergent intensity normalized by its mean value. Due to variations
in the physical conditions, such as temperature contrast, vertical
velocity and overshooting, the intensity contrast varies with stellar
parameters, which is commonly referred as hidden or naked granulation
(see Nordlund \& Dravins \cite{Nordlund:1990p6720}). In Fig. \ref{fig:intensity-contrast},
we show the latter for different stellar parameters. Interestingly,
we find at lower metallicity an enhancement in the covered range of
the intensity contrast. In the illustrated case the enhancement amounts
$60\,\%$. We located the reason for the enhancement being the ionization
of hydrogen, which becomes the main source for the formation of negative
hydrogen, since the metals are depleted.

Furthermore, we derived the granule sizes from the maximum amplitude
of the spatial power spectrum of the emergent intensity. We find the
granule size to scale well with the pressure scale height over our
large range in stellar parameters. The scaling has been suggested
by Stein \& Nordlund \cite{Stein:1998p3801} based on the conservation of mass for the deflecting
mass flux in a simplified picture.

Similar to Ludwig et al. \cite{Ludwig:1999p7606}, we matched the free mixing length
parameter $\alpha_{\mathrm{MLT}}$ with the adiabatic entropy value
of the deep convection zone with a 1D code that uses the identical EOS
and opacity. The mixing length systematically decreases for higher
$\teff$, lower $\logg$ and higher metallicity. The cool dwarfs exhibit
the largest $\alpha_{\mathrm{MLT}}$, which reflects a very large
convective efficiency or similarly a very low entropy jump.

\section{The mean stratification}

\begin{figure}
\includegraphics[width=88mm]{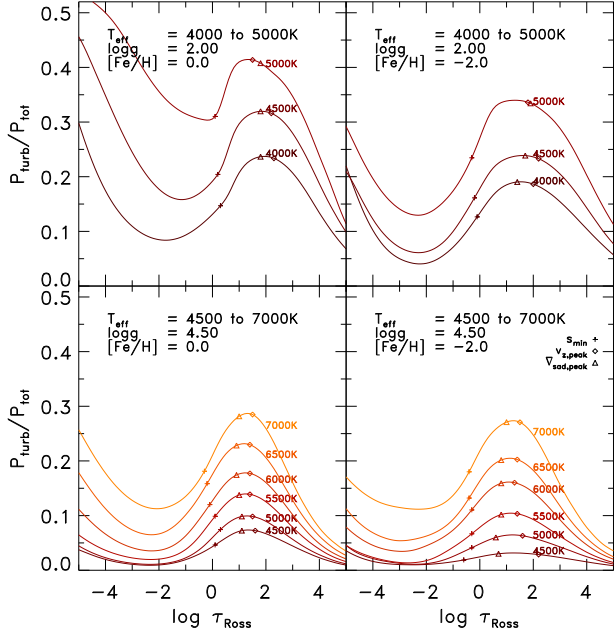}

\caption{\label{fig:pturb}Fraction of the turbulent pressure on the total
pressure.}
\end{figure}
The velocity field in our 3D RHD simulations emerge self-consistently
from the radiative cooling of the hot upflowing granules taking place
at the photospheric transition region, which leads to large entropy
fluctuations that drives convection through buoyancy work of the turbulent
downdrafts. In Fig. \ref{fig:pturb}, we show the fraction of the
turbulent pressure, $p_{\mathrm{turb}}=\rho v_{z}^{2}$, for different
stellar parameters. The specific depth-dependence arises from the
changes of the vertical velocity, $v_{z}$. Below the optical surface
($\ltaur>0$), the vertical velocity reaches its maximum in the superadiabatic
regime. The maximum increases with higher $\teff$, lower $\logg$
and higher $\feh$. We note that 1D models predict rather different
velocities, since these are based on MLT, and serve only for the retrieval
of the convective energy flux. Observed 3D effects like overshooting
and asymmetry in the velocity are rendered astonishingly accurate
in the case of the solar simulation with the 3D models, while 1D models
are incapable of doing so. Furthermore, the turbulent pressure is
often neglected in 1D models, however, as given in Fig. \ref{fig:pturb},
the contribution of the turbulent pressure can be non-negligible,
in particular, for hotter stars or towards giants.

\section{Summary}

We have computed a large grid of realistic 3D atmosphere models. These
exhibit a large amount of properties that has to analyzed carefully,
in order to understand subsurface convection better. We intend to
perform various applications with the grid models in order to improve
the theoretical predictions.


\end{document}